\documentclass{article}

\PassOptionsToPackage{numbers, square}{natbib}
\usepackage[final]{neurips_2025}

\usepackage[utf8]{inputenc} % allow utf-8 input
\usepackage[T1]{fontenc}    % use 8-bit T1 fonts
\usepackage{hyperref}       % hyperlinks
\usepackage{url}            % simple URL typesetting
\usepackage{booktabs}       % professional-quality tables
\usepackage{amsfonts}       % blackboard math symbols
\usepackage{nicefrac}       % compact symbols for 1/2, etc.
\usepackage{microtype}      % microtypography
\usepackage{xcolor}         % colors
\usepackage[pdftex]{graphicx} 
\usepackage{caption}
\usepackage{float} 
\usepackage{subfig}
\usepackage{algorithm}
\usepackage{algpseudocode}
\usepackage{amsmath}
\usepackage{makecell}
\usepackage{multirow}
\usepackage{multicol}
\usepackage{hyperref}

\title{Synergistic Tensor and Pipeline Parallelism}

\author{
    Mengshi Qi\textsuperscript{\thanks{Corresponding author: qms@bupt.edu.cn.} \hspace{0.5mm} 1}, Jiaxuan Peng\textsuperscript{1}, Jie Zhang, Juan Zhu, Yong Li, Huadong Ma\textsuperscript{1} \\
    \textsuperscript{1}State Key Laboratory of Networking and Switching Technology,\\
    Beijing University of Posts and Telecommunications, China\\
    \texttt{\{qms, pjx, mhd\}@bupt.edu.cn}
}

\begin{document}

\maketitle

\begin{abstract}
In the machine learning system, the hybrid model parallelism combining tensor parallelism (TP) and pipeline parallelism (PP) has become the dominant solution for distributed training of Large Language Models~(LLMs) and Multimodal LLMs (MLLMs). However, TP introduces significant collective communication overheads, while PP suffers from synchronization inefficiencies such as pipeline bubbles. Existing works primarily address these challenges from isolated perspectives, focusing either on overlapping TP communication or on flexible PP scheduling to mitigate pipeline bubbles. In this paper, we propose a new synergistic tensor and pipeline parallelism schedule that simultaneously reduces both types of bubbles. Our proposed schedule decouples the forward and backward passes in PP into fine-grained computation units, which are then braided to form a composite computation sequence. This compositional structure enables near-complete elimination of TP-related bubbles. Building upon this structure, we further design the PP schedule to minimize PP bubbles. Experimental results demonstrate that our approach improves training throughput by up to 12\% for LLMs and 16\% for MLLMs compared to existing scheduling methods. Our source code is avaiable at \href{https://github.com/MICLAB-BUPT/STP}{https://github.com/MICLAB-BUPT/STP}.

\end{abstract}

\section{Introduction}
\begin{figure}
    \centering
    \includegraphics[width=0.8\linewidth]{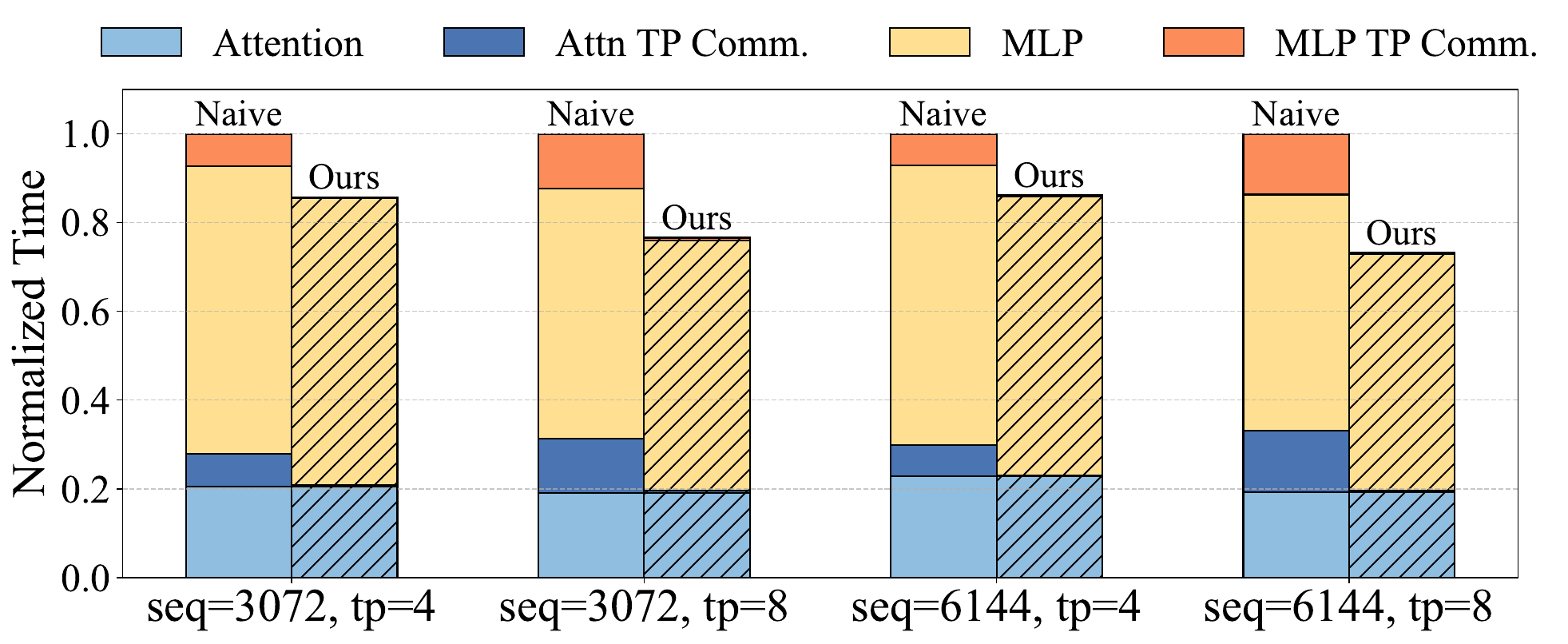}
    % \vspace{-2mm}
    \caption{Speedup of overlapping TP communication with computation in PP within a Transformer layer of Qwen2~\cite{yang2024qwen2technicalreport} in forward pass. The proportion of TP communications grows significantly with the increased TP size, which are effectively overlapped in our schedule compared with the naive implementation.}
    \label{fig: tp comm ratio}
    \vspace{-2mm}
\end{figure}

Distributed systems~\cite{huang2024distmm, narayanan2021efficient, wang2024accelerating, jiang2024dynapipe, rajbhandari2022deepspeed-moe} have become the cornerstone for the training of large-scale machine learning models~\cite{brown2020language, touvron2023llama, qi2025action, deng2025global, lv2023disentangled, qi2019attentive, qi2021semantics, lv2024sgformer}, particularly since model scales have grown to tens of billions of parameters. Data parallelism~\cite{jiang2020unified, li2014communication, goyal2017accurate} is commonly employed to accelerate training and is most effective for relatively smaller models. As the scale of the model increases, researchers have turned to model parallelism~\cite{jia2019beyond, shazeer2018mesh, SDPipe}, which partitions the model into smaller components and distributes them between multiple devices. Model parallelism primarily consists of two strategies: tensor parallelism~(TP)~\cite{shoeybi2019megatron} and pipeline parallelism~(PP)~\cite{huang2019gpipe, narayanan2021efficient, qi2024zero}. Tensor parallelism partitions model weights across devices according to predefined rules, thereby reducing per-device memory requirements. While TP alleviates GPU memory pressure to some extent, it introduces additional collective communication operations within TP groups, referred to as TP bubbles that exhibit significant growth with increasing TP size, as shown in Figure~\ref{fig: tp comm ratio}. For instance, the TP bubbles account for 27.5\% of the overall time in the configuration of TP=8 and sequence length of 6,144, which can significantly degrade the performance. In contrast, pipeline parallelism divides the model into chunks along the layer dimension, which are assigned to PP stages according to the PP strategy. Activations and gradients are then transmitted among PP stages during the forward and backward passes. This strategy enables efficient utilization of computational resources but requires careful synchronization between stages to minimize idle time, commonly referred to as PP bubbles.

% To mitigate TP bubble issues, existing studies~\cite{wang2022overlap, jangda2022breaking, chang2024flux} aim to overlap the communication with computation at the kernel~\cite{chang2024flux} or hardware levels~\cite{pati2024t3}.
To mitigate TP bubble issues, existing studies~\cite{wang2022overlap, jangda2022breaking, chang2024flux, pati2024t3} aim to overlap communication with computation at the kernel level, such as customized CUDA kernels~\cite{chang2024flux}, or at the hardware level, where specialized hardware mechanisms like the track-and-trigger system~\cite{pati2024t3} are employed.
By carefully scheduling paired decomposed communication and computation chunks, these methods effectively reduce overall execution time. As for PP bubbles, GPipe~\cite{huang2019gpipe} reduces pipeline bubbles by splitting the global batch into microbatches. Subsequently, 1F1B~\cite{narayanan2019pipedream} improves upon GPipe by scheduling backward passes earlier, thereby optimizing peak memory footprint. Furthermore, 1F1B-I~\cite{narayanan2021efficient} reduces the bubble rate by introducing virtual stages assigned to individual devices. More recently, Zero Bubble~\cite{qi2024zero} decouples the full backward pass into separate activation and weight gradient computations, enabling more flexible scheduling and achieving a significantly lower PP bubble rate.

% As analyzed before, the key challenge in TP and PP lies in the inherent communication overheads introduced by TP, along with the synchronization overheads in PP.
In practice, hybrid model parallelism, particularly the combination of PP and TP, has become the dominant approach for large-scale model training. However, most of the aforementioned works primarily focus on optimization from isolated perspectives, generally yielding limited enhancements. Inspired by studies~\cite{liu2024deepseek, korthikanti2023reducing} that integrate two kinds of parallelism strategies for better optimization, our work aims to integrate PP and TP from a synergistic perspective. Specifically, our method leverages the abundant computation operations inherent in PP to overlap the TP bubbles and employs precise scheduling to mitigate the PP bubbles. Firstly, we decouple the forward and backward passes of the model chunk within each stage of PP into fine-grained computation units. By interleaving forward and backward computation units to construct execution blocks, we achieve near-zero TP idle time as shown in Figure~\ref{fig: tp comm ratio}, where TP communication and the fine-grained computation units in PP execute in parallel. Building upon these blocks, we further design a PP schedule with a low bubble rate and balanced memory footprints across stages. Additionally, we provide an enhanced variant of our schedule that incorporates activation offloading, serving as a preliminary attempt under limited memory conditions. Our main contributions are summarized as follows:
\begin{itemize}
    \item We rethink hybrid model parallelism from a synergistic perspective and decouple the forward and backward passes of PP into fine-grained computation units, which are interleaved together to form braided execution blocks that effectively eliminate TP communication bubbles at the scheduling level.
    \item We design a novel PP schedule that synergistically integrates with TP based on the braided execution blocks. This schedule features a ``V''-shape dataflow that achieves balanced memory footprints and a significantly reduced PP bubble rate.
    \item We conduct extensive experiments on popular LLMs and MLLMs of various scales to validate the universality and effectiveness of our method. The results demonstrate that our proposed schedule improves throughput by up to 16\% compared to the existing methods.
\end{itemize}

% These overheads are unavoidable in current scheduling strategies, directly increasing execution time and degrading end-to-end performance, particularly in multimodal LLMs, which comprise more layers than standard LLMs. This issue becomes particularly severe in high-TP configurations for large-scale models, where the communication-computation ratio exhibits exponential growth with increasing TP size, as demonstrated in Figure~\ref{fig: tp comm ratio}. Moreover, the decoupled backward pass mechanism in Zero Bubble disrupts the communication-computation overlap during the backward phase, resulting in higher TP overhead compared to alternative scheduling approaches.

\section{Related Work}
\paragraph{Tensor Parallelism.} Tensor parallelism~\cite{shoeybi2019megatron, narayanan2021efficient, korthikanti2023reducing, wang2022overlap, jangda2022breaking, chang2024flux} partitions tensors across devices to reduce per-device memory usage but introduces communication overhead due to synchronization requirements. This problem becomes more pronounced with TP sizes increase, where the limited inter-device bandwidth (\emph{e.g.}, systems without NVLink~\cite{nvidia2025nvlink}) can severely degrade training efficiency. Several prior works~\cite{wang2022overlap, jangda2022breaking, chang2024flux} optimize this kind of collective communication by breaking down both communication and corresponding computation operations into sequences of smaller chunks. Subsequently, they carefully coordinate the execution of these granular units to achieve acceleration, often through the intricate fusion of computation and communication kernels, along with complex hardware control. In contrast, our proposed schedule optimizes TP communication at the scheduling or software level, providing a more accessible and user-friendly approach.

\paragraph{Pipeline Parallelism.} Pipeline parallelism~\cite{chimera, jiang2024dynapipe, AdaPipe, jeon2025graphpipe, liu2023hanayo, kim2023bpipe, liu2024deepseek} divides the entire model into smaller chunks using a variety of strategies. GPipe~\cite{huang2019gpipe} reduces PP bubbles by splitting the global batch into smaller micro-batches. Building on this, PipeDream~\cite{narayanan2019pipedream} improves upon GPipe by advancing the backward computation to further reduce memory usage. 1F1B-I~\cite{narayanan2021efficient} and Hanayo~\cite{liu2023hanayo} further introduce additional virtual stages to lower the bubble rate. More recently, Zero Bubble~\cite{qi2024zero} decouples the backward computation into activation gradient and weight gradient computation to achieve a near-zero PP bubble schedule. However, these existing schedules tend to address PP bubbles in isolation. In this paper, we design an integrated approach that concurrently minimizes both TP and PP bubbles by synergistically combining these two parallelism techniques.

\begin{figure}[!ht]
    \centering
    \includegraphics[width=0.9\linewidth]{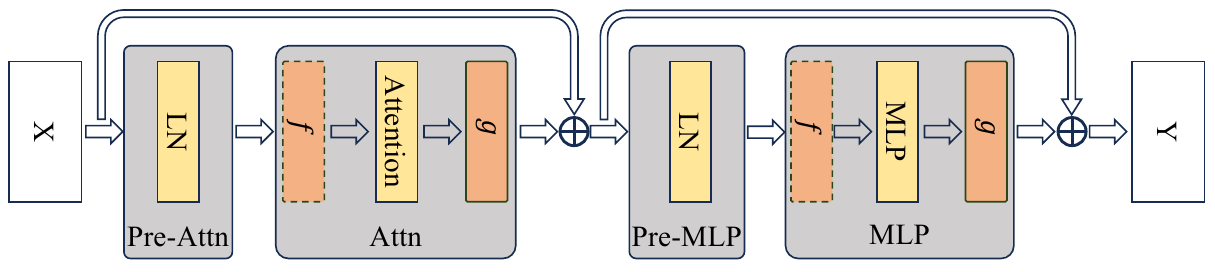}
    \vspace{-1mm}
    \caption{Illustration of the main computation and communication operation in a single TP Transformer layer. $f$ is an identity operation in forward while All-Reduce in backward. $g$ is opposite to $f$.}
    \label{tp layer fig}
    %%\vspace{-1mm}
\end{figure}

\begin{figure}[!ht]
    \centering
    \subfloat[Execution block with forward and full backward.]{\includegraphics[width=\linewidth]{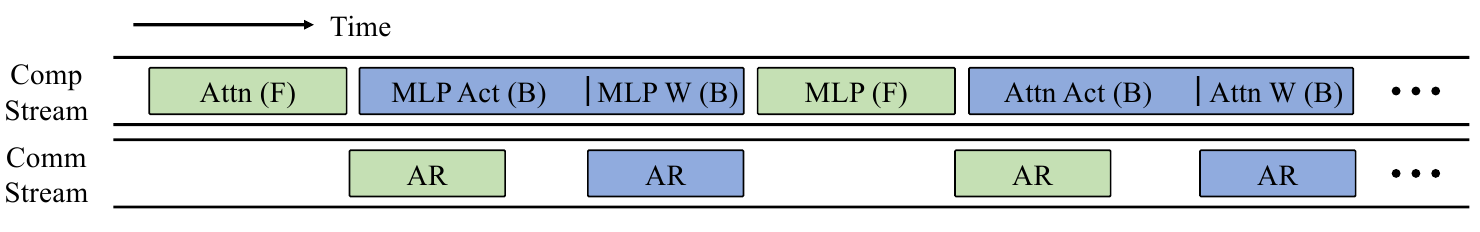} \label{overlap 1}}
    \newline
    \vspace{-2mm}
    \subfloat[Execution block with forward and activation backward.]{\includegraphics[width=\linewidth]{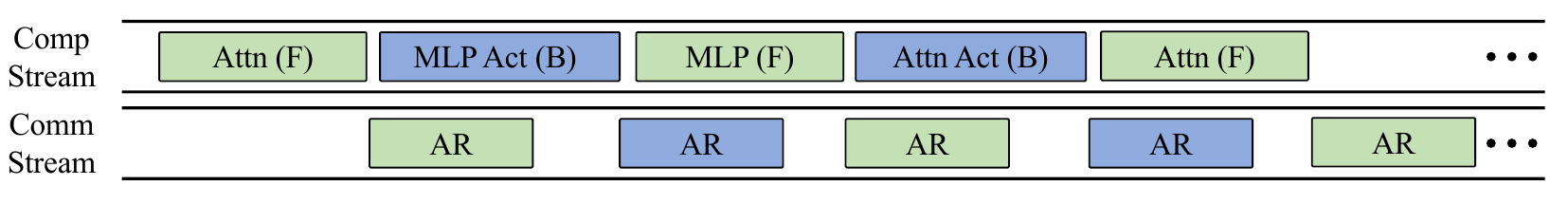} \label{overlap 2}}
    \caption{Two types of execution blocks that braid the forward (F) and backward (B) computation units to overlap the TP communications (All-Reduce), denoted as AR.}
    \label{overlapped fig}
    \vspace{-1mm}
\end{figure}

\section{Braided Execution Block}
\label{sec:3}
Although TP reduces memory consumption, it introduces additional communication overhead after the Attention and MLP computations in both the forward and backward passes, as illustrated in the orange blocks in Figure~\ref{tp layer fig}. As shown in the blue blocks of Figure~\ref{overlap 1}, the communication represented by the blocks in the communication stream can be naturally overlapped with the weight gradient computation during the backward pass,  where TP communication (All-Reduce) and weight gradient computation run in parallel. However, the communication in the solid blocks during the forward pass cannot be easily overlapped with other operations at the software level due to data dependencies, leading to massive TP bubbles. To address this TP bubble issue, we design two types of execution blocks that overlap TP communication with PP computation through braiding forward and backward computation units, as shown in Figure~\ref{overlapped fig}. These blocks serve as the foundation of our schedule.

Specifically, we decompose the entire Transformer layer into several fine-grained computation units: the Pre-Attn unit, Attn unit, Pre-MLP unit, and MLP unit. The units from the forward and backward passes are then interleaved sequentially to form a braided execution block, as shown in Figure~\ref{overlap 1}. Moreover, both the Attn and MLP units in backward passes are further split into two components: activation gradient computation and weight gradient computation, following Zero Bubble~\cite{qi2024zero}. When necessary, we activate the separation of weight gradient computations to balance the workload and reduce pipeline bubbles. This separation does not disrupt the execution blocks, as the subsequent forward computation can fill the bubbles introduced by the separation, as shown in Figure~\ref{overlap 2}. The Pre-Attn and Pre-MLP units are inserted into the computation stream according to their computational dependencies. All these operations only involve model-level modifications, making our approach more user-friendly compared to kernel-level implementations.

Furthermore, considering the residual computations after the Attn and MLP units that introduce additional data dependencies and increase engineering complexity for backward passes, we modify the forward and backward computations to fuse the residual computation into the Attn and MLP units before applying operation $g$. This modification preserves computational equivalence and does not affect model convergence. The modified forward computation of the Attn unit for a single TP rank can be formulated as follows:
\begin{equation}
    X_{ln} = LayerNorm(X), \quad X_{attn} = AR(Attention(X_{ln}) + \frac{detach(X)}{t}),
\end{equation}

where $t$ denotes the tensor parallel size, $detach(\cdot)$ is an operation that stops gradient computation of $X$, $Attention(\cdot)$ is the attention computation performed without any communication, and $AR(\cdot)$ represents the All-Reduce communication operation. The gradient computation $g(X)$ of the hidden state $X$ is then formulated as the following:
\begin{equation}
    g(X) = AR\left(\frac{\partial X_{attn}}{\partial X_{ln}}\right) \cdot \frac{\partial X_{ln}}{\partial X} + 1,
\end{equation}
where the term $+1$ accounts for the gradient contribution from the residual connection. Similarly, the MLP unit adopts a comparable computation fusion strategy.

\section{Pipeline Parallelism Schedule}
We begin by introducing several key principles that facilitate the construction of the overall schedule in Section~\ref{sec:4.1} and then introduce the proposed schedule in Section~\ref{sec:sch}. Following this, we present a theoretical analysis of the PP and TP bubbles and peak memory in Section~\ref{sec:4.2}. Finally, we propose an enhanced variant of our schedule to accommodate scenarios with limited memory in Section~\ref{sec:4.3}.

\subsection{Building Principles}
\label{sec:4.1}

To ensure an effective pipeline schedule, we outline several guiding principles aimed at optimizing memory efficiency and minimizing bubbles.

\begin{figure}[!ht]
    \centering
    \includegraphics[width=\linewidth]{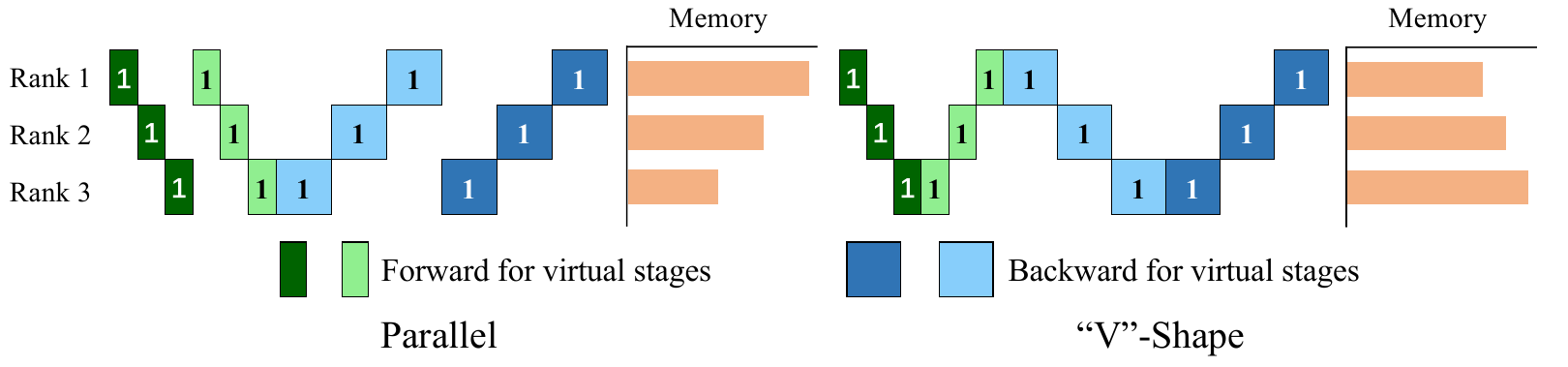}
    \caption{Comparison between parallel and ``V''-shape pipeline flow for microbatch 1, which shows improved memory balance across stages. That is attributed to the early backward pass on device 0.}
    \label{fig: basic block}
    \vspace{-2mm}
\end{figure}

\paragraph{Balanced Memory Footprint:} As discussed in~\cite{qi2024pipeline, liu2023hanayo}, the 1F1B-I~\cite{narayanan2021efficient} often leads to a memory bottleneck on the first device due to the parallel dataflow of virtual stages, while subsequent stages often have underutilized memory resources. This imbalance results in inefficient resource usage. To tackle this issue, current scheduling methods~\cite{liu2023hanayo, qi2024pipeline, liu2024deepseek} utilize a ``V''-shape scheduling strategy to achieve a more balanced distribution of peak memory across all devices (see Figure~\ref{fig: basic block}). We follow this principle by implementing a "V"-shape dataflow for our virtual stages to optimize memory efficiency, which also supports the scheduling of braided execution blocks for improved overall performance.

\paragraph{Bubble Minimization:} The braided computation-communication execution blocks, as detailed in Section~\ref{sec:3}, illustrate the arrangement of computation units. The practical construction of these blocks follows two primary patterns: (1) pairing the forward units of one virtual stage with the backward units of another, or (2) braiding the forward and backward units of the same model chunk across different microbatches. The selection of the building pattern is closely tied to the characteristics of virtual stages in PP. Specifically, when partitioning models into virtual stages, it is essential to maintain approximately equal execution times across these stages to mitigate pipeline bubbles. For MLLMs, a significant disparity often exists between the vision encoder (\emph{e.g.}, ViT) and the Language Model~(LM) due to differences in their hidden sizes. This presents a key challenge for multimodal models, where earlier virtual stages typically contain more ViT layers, while later stages consist of fewer LM layers to achieve balanced execution times. However, the imbalance in the number of computation units across virtual stages limits the effectiveness of pattern (1), as mismatched unit counts prevent efficient hiding of TP communication during cross-virtual-stage pairings. In contrast, pattern (2), which involves overlapping the forward and backward units of the same chunk, more effectively minimizes the TP bubbles. Therefore,  we adopt pattern (2) to construct the basic blocks for universality, especially considering the multimodal models.

\begin{figure}[!ht]
    \centering
    \includegraphics[width=\linewidth]{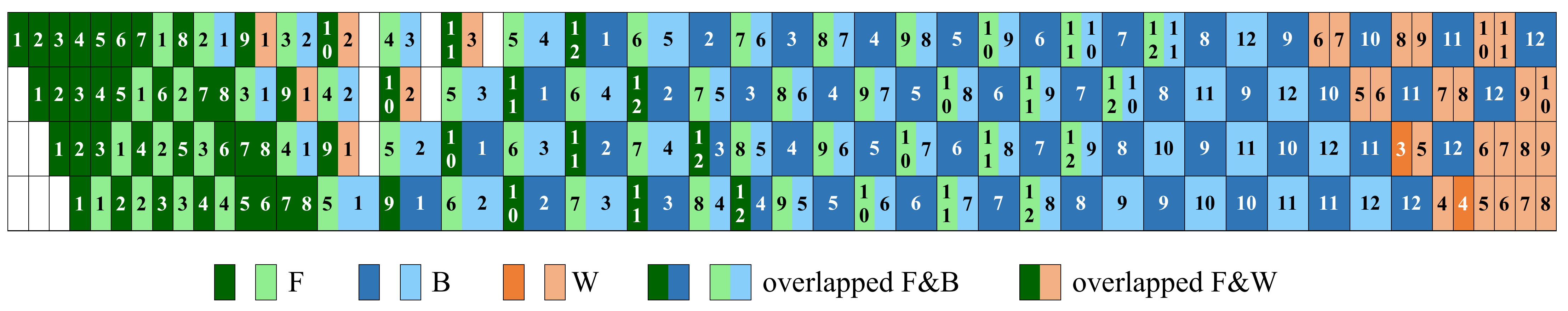}
    % \vspace{-5mm}
    \caption{Synergistic tensor and pipeline parallel schedule with the setting of 4 devices and 12 microbatches. The dark and light pieces indicate the computation of model chunks 0 and 1, respectively. F, B, and W represent the forward, backward, and backward for weight gradients, respectively.}
    \label{schedule}
    \vspace{-2mm}
\end{figure}

\subsection{Synergistic Tensor and Pipeline Parallel Schedule}
\label{sec:sch}
Based on the above building principles, we construct a synergistic tensor and pipeline schedule, which implements a ``V''-shape dataflow across stages for a single microbatch, as illustrated in Figure~\ref{schedule}. Similar to existing schedules~\cite{narayanan2019pipedream, narayanan2021efficient, qi2024zero}, our proposed method comprises three main phases: the warm-up phase, the steady phase, and the cool-down phase. First, the warm-up phase begins with the maximum feasible number of in-flight microbatches before the commencement of the first backward pass. This helps to maintain a low PP rate and facilitates the construction of the braided blocks, as the backward passes of earlier microbatches form the blocks in conjunction with the forward passes of subsequent microbatches. The number of microbatches involved in the warm-up phase is determined by the number of stages. During this phase, the overlapped forward and backward (F\&B) start from the braided computation of the first and second microbatches, and weight gradient separation is activated to quickly propagate gradients to the next stage, except for the last stage.

% v1
After several iterations of overlapped forward and weight (F\&W) and F\&B, the schedule transitions into the steady phase. In this phase, weight separation is deactivated as long as new microbatches continue to arrive. Under these conditions, the system performs one F\&B for model chunk 1, followed by one F\&B for chunk 0. When the supply of microbatches is exhausted, maintaining full F\&B becomes infeasible, and the process degrades into a full backward pass followed by a separated F\&B. During this degraded phase, weight separation is reactivated to align the time of F\&B with that of the full backward pass. During the cool-down phase, the backward passes for all remaining in-flight microbatches are completed. The pipeline bubbles are filled with stored weight gradient computations from earlier steps. Moreover, activating weight separation during the steady phase leads to memory accumulation, as the corresponding weight computations remain pending until the final steps. Therefore, the saved activations required for the weight gradients are offloaded to the CPU in parallel with the computation streams and reloaded when necessary.

\subsection{Theoretical Analysis}
\label{sec:4.2}

We define the number of PP stages and microbatches as $p$ and $m$ respectively, where $p \ll m$. For simplicity, multimodal models are not taken into account in this context. For LLMs, we assume the model is evenly divided into chunks along with the layer dimension, and each microbatch requires the same activation memory $M_a$ for each model chunk across all devices. The computation time for the forward pass of one chunk is denoted as $T_F$, while the durations of activation and weight gradient computation are $T_B$ and $T_W$, respectively. The TP communication time of one model chunk is represented as $T_{AR}$, which remains consistent for both forward and backward passes.

\begin{table}[!htbp]
    \vspace{-2mm}
    \caption{Comparison between 1F1B-I, Zero Bubble V, and our proposed schedule on theoretical PP bubble size, non-overlapped TP communication bubble, and peak activation memory.}
    \vspace{2mm}
    \label{tab:memory}
    \centering
    \begin{tabular}{cccc}
         \toprule
         Schedule & PP Bubble & TP Bubble & Peak Act. Memory\\
         \midrule
         1F1B-I & $(p-1)*(T_{F}+T_{AR} + T_B+T_W)$ & $2*m*T_{AR}$ & $(3*p-2) * M_a$\\
         ZB-V & $(p-1)*(T_{F}+2T_{AR}+ T_B-2T_W)$ & $4*m*T_{AR}$ & $2*p * M_a$ \\
         Ours & $(p-1)*(T_{F}+T_{AR}+T_B-T_W)$ & $(2*p+1)*T_{AR}$ & $3* p * M_a$ \\
         \bottomrule
    \end{tabular}
    % \vspace{-3mm}
\end{table}

As shown in Table~\ref{tab:memory}, 1F1B-I~\cite{narayanan2021efficient} with 2 virtual stages requires $(3*p-2) * M_a$ peak activation memory for the first device, which progressively decreases as the stage index increases. In the backward pass, the TP communications are overlapped naturally, but the TP communications in the forward pass are non-overlapped, resulting in a total TP communication time is $2*m*T_{AR}$. In contrast, Zero Bubble V (ZB-V)~\cite{qi2024pipeline} achieves a smaller theoretical bubble size of $(p-1)*(T_{F}+2T_{AR}+ T_B-2T_W)$ and maintains the lower peak activation memory than that of 1F1B-I. However, due to the decoupling of the backward into activation and weight gradient computations, the TP communication in both forward and activation backward pass becomes non-overlapped, leading to a total TP communication time of $4*m*T_{AR}$ and a larger practical bubble size. Our schedule exhibits the approximate bubble size of $(p-1)*(T_{F}+T_{AR}+T_B-T_W)$ compared to the ZB-V schedule, which is much smaller than that of 1F1B-I. And it achieves substantially reduced TP overheads, benefiting from the overlapped computation and communication, albeit with a trade-off in the peak activation memory.

\subsection{Enhanced Pipeline Schedule}
\label{sec:4.3}
Considering the practical demands of limited CUDA memory, we attempt to reduce the peak memory footprint of our scheduling strategy. The elevated peak memory usage primarily stems from two phases: the warm-up phase, where the number of in-flight microbatches is maximized to maintain a low bubble rate, and the steady phase, where activations from model chunk 0 persist for an extended period before being released, denoted as lifespan. Existing methods~\cite{yuan2024accelerating, AdaPipe} such as activation checkpoint and offloading have been used to reduce memory. However, checkpointing introduces an extra forward pass during backward, increasing execution time. Thus, activation offloading offers a better trade-off between memory reduction and e2e performance, when high-speed bandwidth is available between the host and the device.

\begin{figure}[!htbp]
    \centering
    \includegraphics[width=0.95\linewidth]{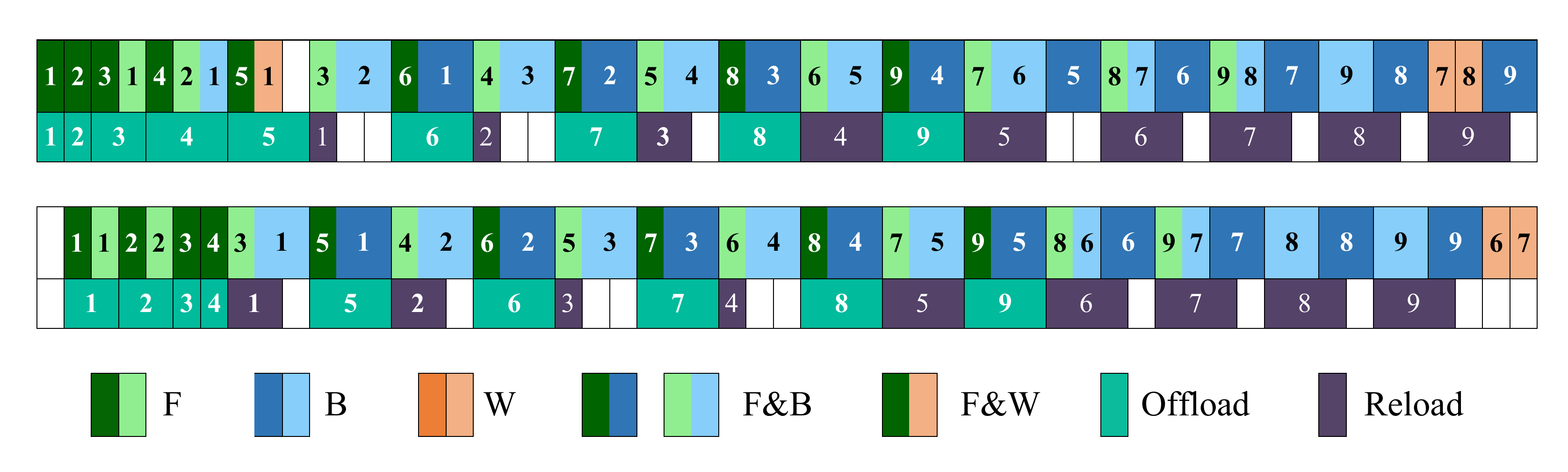}
    \vspace{-2mm}
    \caption{Illustration of our proposed schedule with offloading and reloading for one microbatch in warm-up and steady phase with the setting of PP size 2.}
    \label{fig: offload}
    \vspace{-1mm}
\end{figure}

During the warm-up phase in Figure~\ref{fig: offload}, where memory usage accumulates rapidly, the activations of chunk 1 have a short lifespan and are thus deprioritized for offload. The offload time $T_o$, determined by the offloading ratio $\alpha$, is restricted to be less than the forward time $T_F$ to avoid interference with subsequent operations. In the steady phase, chunk 0 activations persist much longer than those in chunk 1, making them the primary target for offloading. The offload ratio $\alpha$ in this phase can be set higher than in the warm-up phase due to the braided execution blocks. To minimize PCIe bandwidth contention caused by frequent dual CPU-GPU data transfers, we avoid offloading chunk 1 activations, ensuring they can be reloaded in time without disrupting computation. The ratio $\alpha$ is adjusted based on hardware characteristics such as PCIe bandwidth and FLOPs.

\section{Experiments}
\subsection{Setup}

We evaluated our proposed schedule on the series of Qwen2 (LLM)~\cite{yang2024qwen2technicalreport} and Qwen2-VL (MLLM)~\cite{wang2024qwen2vlenhancingvisionlanguagemodels} models, as detailed in Table~\ref{tab:model config}. Meanwhile, Flash Attention 2~\cite{dao2023flashattention} is leveraged in all models for efficiency. Our implementation is built upon the open-source Megatron-Core project~\cite{shoeybi2019megatron} and tested on up to 32 NVIDIA A800 SXM4 80G GPUs distributed across 4 nodes. Each experimental result is recorded after several warm-up iterations to ensure stability. 

In LLM scenarios, we adopt the approach described in~\cite{qi2024zero}, uniformly splitting the model while ensuring that the last stage contains two fewer layers than the other stages, considering the large vocabulary size of 152,064 in Qwen. In MLLM scenarios, the ViT encoder is assigned to the first virtual stage on device 0, and the LM model is uniformly distributed across the remaining virtual stages, and the last virtual stage also contains two fewer layers compared to the other stages.

To be specific, our experiments primarily concentrate on the following pipeline parallel schedules: a) interleaved 1F1B (1F1B-I), as implemented in Megatron-LM~\cite{narayanan2021efficient}, which reduces the bubble ratio by interleaving stages among workers but slightly increases memory footprint; b) ZB-V, introduced in~\cite{qi2024pipeline}, which achieves a balanced peak memory distribution through a ``V''-shape dataflow; and c) our proposed schedule. For consistency, all schedules are configured with two virtual stages per device, and the implementations are based on open-source codebases.

\vspace{-2mm}
\begin{table}[!h]
    \centering
    \caption{Model configurations evaluated on the experiments.}
    \vspace{2mm}
    \label{tab:model config}
    \begin{tabular}{ccccccccc}
    \toprule
     Model & Scale & \makecell[c]{ViT \\ Layers} & \makecell[c]{ViT \\ Heads} & \makecell[c]{ViT \\ Dim.} & \makecell[c]{LM \\ Layers} & \makecell[c]{LM \\ Q Heads} & \makecell[c]{LM \\ KV Heads} &  \makecell[c]{LM \\ Dim.} \\
    \midrule
     LLM & 12.1B & - & - & - & 30 & 40 & 8 & 5120 \\
     LLM & 26.3B & - & - & - & 46 & 56 & 8 & 7168 \\
     MLLM & 14.9B & 32 & 16 & 2048 & 33 & 40 & 8 & 5120 \\
     MLLM &28.8B+ & 26 & 16 & 4096 & \makecell[c]{40/43} & 56 & 8 & 7168 \\
    \bottomrule
    \end{tabular}
    \vspace{-2mm}
\end{table}

\begin{figure}[!h]
    \vspace{-1mm}
    \centering
    \includegraphics[width=\linewidth]{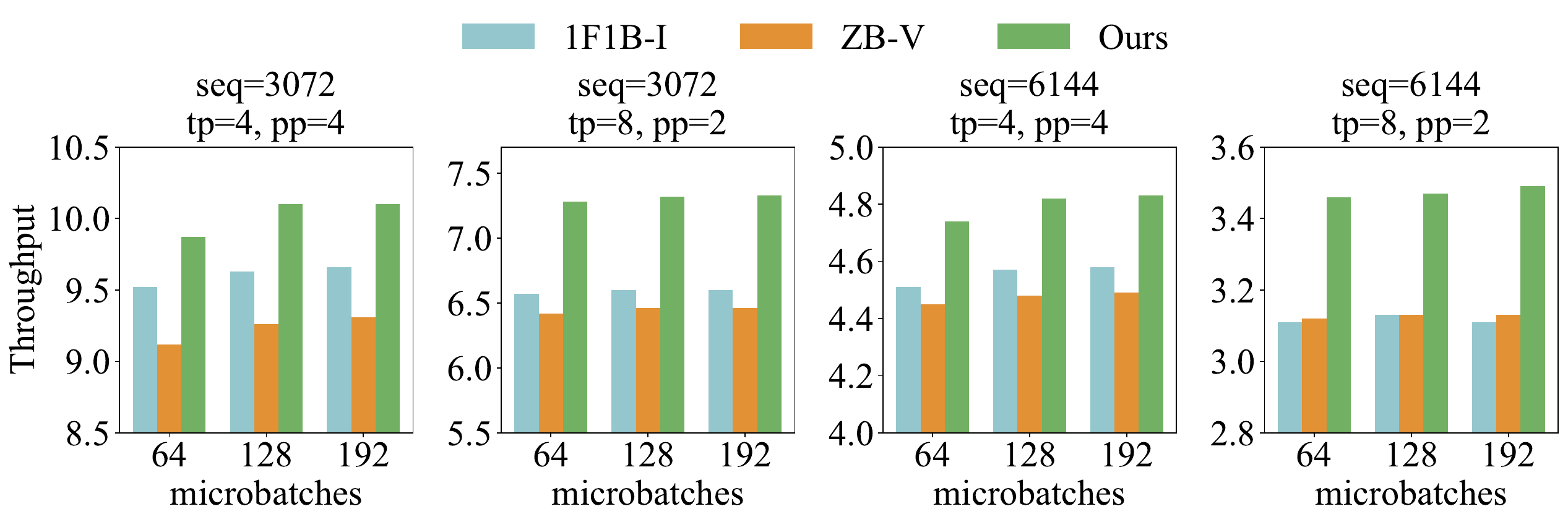}
    \caption{Experiment results on 12.1B LLM across 16 GPUs in terms of throughput (samples per second).}
    \label{fig:lm 2node}
    \vspace{-3mm}
\end{figure}

\begin{figure}[!h]
    \centering
    \includegraphics[width=\linewidth]{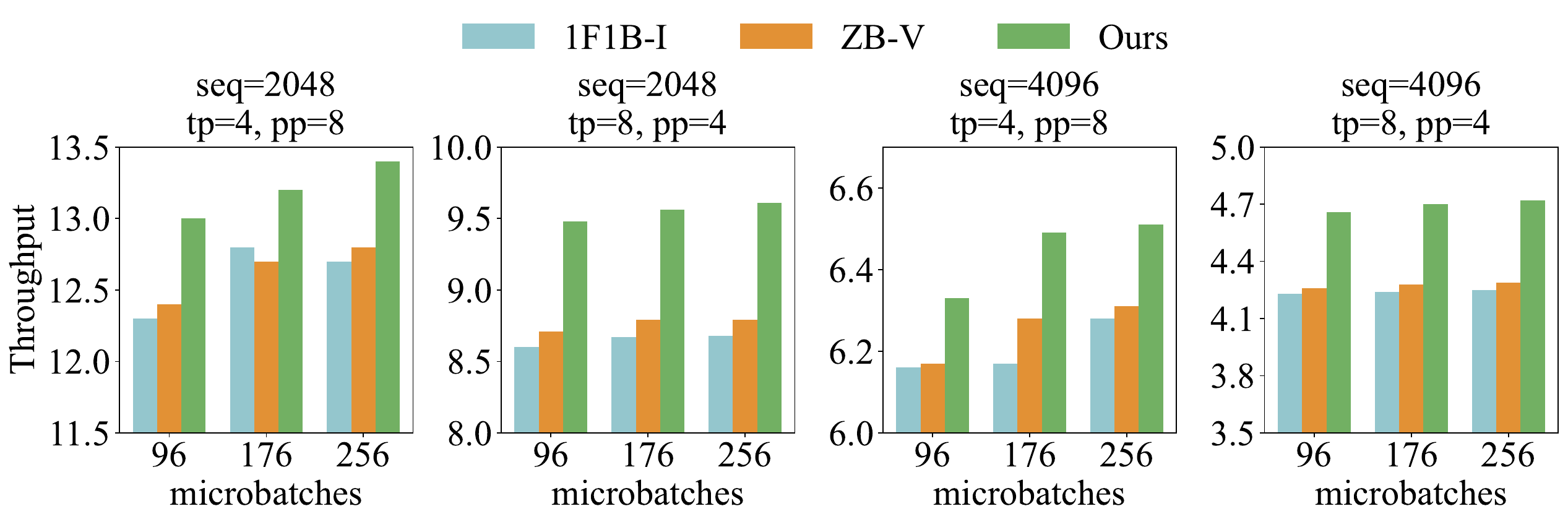}
    \caption{Experiment results on 26.3B LLM across 32 GPUs in terms of throughput. }
    \label{fig:lm 4node}
    \vspace{-3mm}
\end{figure}

\subsection{Comparison with Existing Schedules on LLM}

In Figures~\ref{fig:lm 2node} and~\ref{fig:lm 4node}, we report experimental results by evaluating Qwen2 on 2-node and 4-node setups, respectively, showing the achieved throughput under various pipeline strategies, sequence lengths, and model scale configurations. Our proposed schedule outperforms all other strategies across all settings in terms of throughput. As shown in the right part of Figure~\ref{fig:lm 2node}, our method achieves a maximum improvement of 12.2\% over 1F1B-I. This performance gain is attributed to the large TP size of 8, the long sequence length of 6,144, and the low PP size of 2 used in the experiments. As illustrated in Figure~\ref{fig: tp comm ratio}, TP-related bubbles occupy a significant proportion under TP size 8, especially for attention units, which are effectively mitigated by our scheduling approach.

However, the performance improvements under PP=8 in Figure~\ref{fig:lm 4node} are less pronounced. This is due to the explicit pipeline communication in our schedule, which is executed immediately after computation and cannot be overlapped with computation, thereby negatively impacting performance. Nevertheless, our scheduling strategy still outperforms the other two methods. The highest throughput improvements for both 12B and 26B LLMs are achieved at TP=8, demonstrating that our method offers substantial potential when scaling to larger models with TP=8, even TP=16. Notably, ZB-V demonstrates comparable or even worse performance than 1F1B-I. Since ZB-V is based on the decoupling of the full backward pass, which exposes previously overlapped TP bubbles during the full backward in 1F1B-I and introduces additional bubble overheads, thereby eliminating the potential benefit of reduced PP bubbles from decoupling. 

\begin{figure}[!h]
    \centering
    \includegraphics[width=0.7\linewidth]{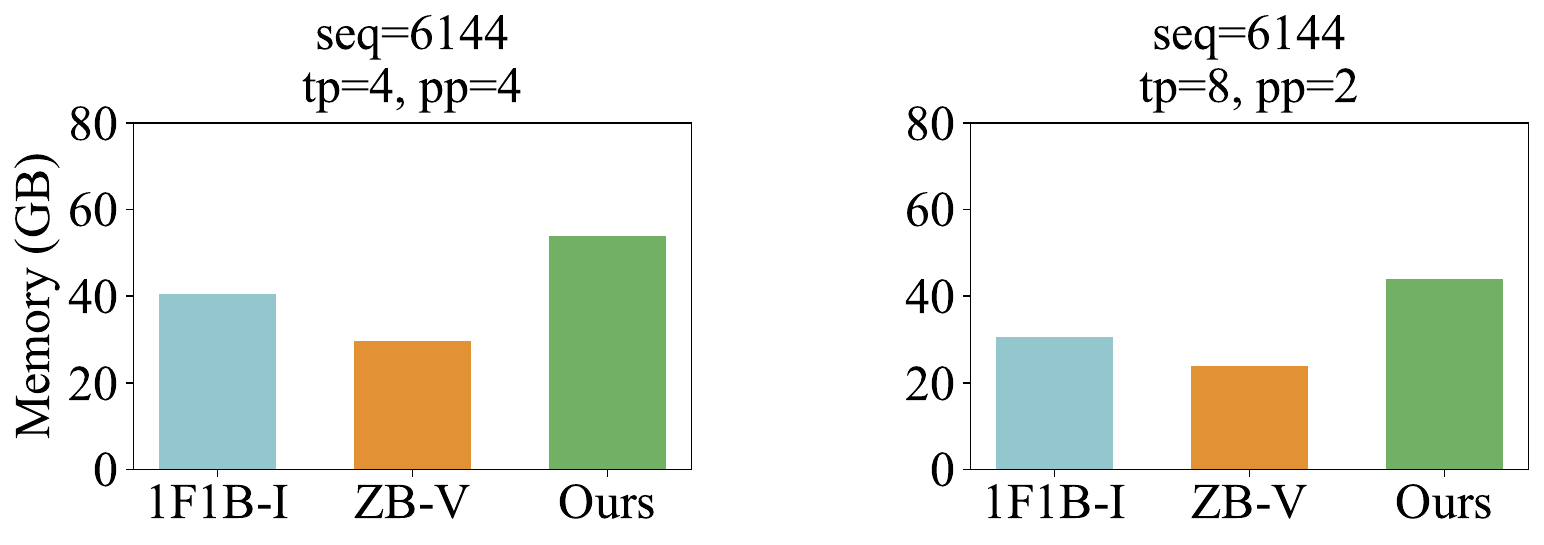}
    %%\vspace{-2mm}
    \caption{Peak activation memory footprint tested on 12.1B LLM across 2 nodes.}
    \label{fig:memory 2 node}
    \vspace{-3mm}
\end{figure}

Furthermore, we show the peak activation memory footprints with 4 and 2 PP stages in Figure~\ref{fig:memory 2 node}. While our schedule exhibits a slightly higher peak memory footprint, it can achieve optimal throughput, especially under memory-unconstrained scenarios. For memory-limited scenarios, we introduce an enhanced variant that significantly reduces memory footprints while maintaining comparable throughput, detailed in Section~\ref{expe}.

% Please refer to the supplementary materials for detailed results.

\begin{table}[!h]
%%\vspace{-3mm}
\centering
\caption{Experimental results for the 14.9B, 28.8B, and 30.3B MLLMs, showing throughput (samples per
second) and peak activation memory footprint (GB). $mbs$ denotes the number of microbatches.}
\vspace{1mm}
\label{tab:mllm 2nodes}
\begin{tabular}{cccccccccc}
\toprule
Model & \makecell[c]{ViT \\ Length}& \makecell[c]{LM \\ Length} & TP & PP & Schedule & \makecell[c]{\\64} & \makecell[c]{$mbs$\\128} & \makecell[c]{\\192} & \makecell[c]{Memory } \\
\midrule
\multirow{6.5}{*}{\makecell[c]{1.7B ViT \\ \\ 13.2B LM\\  \\16 GPUs}} & \multirow{6.5}{*}{3136} & \multirow{6.5}{*}{5120} & \multirow{3}{*}{4} &  \multirow{3}{*}{4} & 1F1B-I & 4.36 & 4.44 & 4.46 & 40 \\
 &  & & & & ZB-V & 4.25 & 4.30 & 4.31 & \textbf{30} \\
 & & & & & Ours & \textbf{4.53} & \textbf{4.63} & \textbf{4.65} & 56 \\
 \cmidrule{4-10}
 & & & \multirow{3}{*}{8} &  \multirow{3}{*}{2} & 1F1B-I & 2.42 & 2.44 & 2.46 & \textbf{26} \\
 & & & & & ZB-V & 2.48 & 2.49 & 2.49 & 30 \\
 & & & & & Ours & \textbf{2.86} & \textbf{2.86} & \textbf{2.87} & 51 \\

\bottomrule
\toprule
Model & \makecell[c]{ViT \\ Length}& \makecell[c]{LM \\ Length} & TP & PP & Schedule & \makecell[c]{\\96} & \makecell[c]{$mbs$\\176} & \makecell[c]{\\256} & \makecell[c]{Memory } \\
\midrule
\multirow{6.5}{*}{\makecell[c]{5.6B ViT \\ \\ 23.2B $/$ \\ 24.7B LM  \\ \\32 GPUs}} & \multirow{3}{*}{1600} & \multirow{3}{*}{3072} & \multirow{3}{*}{4} &  \multirow{3}{*}{8} & 1F1B-I & 5.79 & 5.89 & 5.85 & 60 \\
 & & & & & ZB-V & 5.90 & 5.97 & 6.01 & \textbf{48} \\
 & & & & & Ours & \textbf{5.97} & \textbf{6.19} & \textbf{6.19} & 69 \\
 \cmidrule{2-10}
 & \multirow{3}{*}{2304} & \multirow{3}{*}{5120} & \multirow{3}{*}{8} &  \multirow{3}{*}{4} & 1F1B-I & 3.37 & 3.40 & 3.40 & 43 \\
 & & & & & ZB-V & 3.51 & 3.53 & 3.53 & \textbf{38} \\
 & & & & & Ours & \textbf{3.72} & \textbf{3.80} & \textbf{3.80} & 47 \\

\bottomrule
\end{tabular}
%%\vspace{-3mm}
\end{table}

\subsection{Comparison with Existing Schedules on MLLM}
As shown in Tables~\ref{tab:mllm 2nodes}, we present experimental evaluations of Qwen2-VL on 2-node and 4-node setups, respectively, which report throughput and peak activation memory under various configurations, including ViT lengths, LM lengths, and model scale parameters. For the PP=4, we carefully adjust the ViT and LM sequence lengths to balance computational workload across virtual stages by ensuring the FLOPs of the ViT approximately match those of individual virtual stages. In contrast, for PP=2 and PP=8, we intentionally maintain disparities in ViT/LM FLOPs to conduct diverse workload scenarios. Our scheduling approach demonstrates superior performance compared to baseline methods in all balanced computational scenarios. 
Notably, we observe a maximum throughput improvement of 11.7\% in the configuration with TP=8 and PP=4, which is significantly higher than the 2\% improvement in TP=4 and PP=4, compared to 1F1B-I. This trend aligns with observations from LM experiments, indicating that larger TP sizes generally yield greater performance gains.

A particularly notable result is the 16.7\% performance improvement achieved by our schedule in the PP=2 setting, where the ViT component has relatively lower computational intensity compared to other model chunks. In this case, the lowest peak memory footprint is achieved by 1F1B-I in stage 1, where only a few microbatches are in-flight and the memory in ViT is much smaller compared to that in LM. In the PP=8 scenario, where the ViT's FLOPs exceed those of the LM chunks, our schedule and ZB-V achieve comparable throughputs at $mbs = 96$. However, as the number of microbatches increases, our approach demonstrates superior scalability and robustness, gaining more improvements compared to ZB-V and 1F1B-I.

\begin{figure}[!h]
    \centering
    \includegraphics[width=0.9\linewidth]{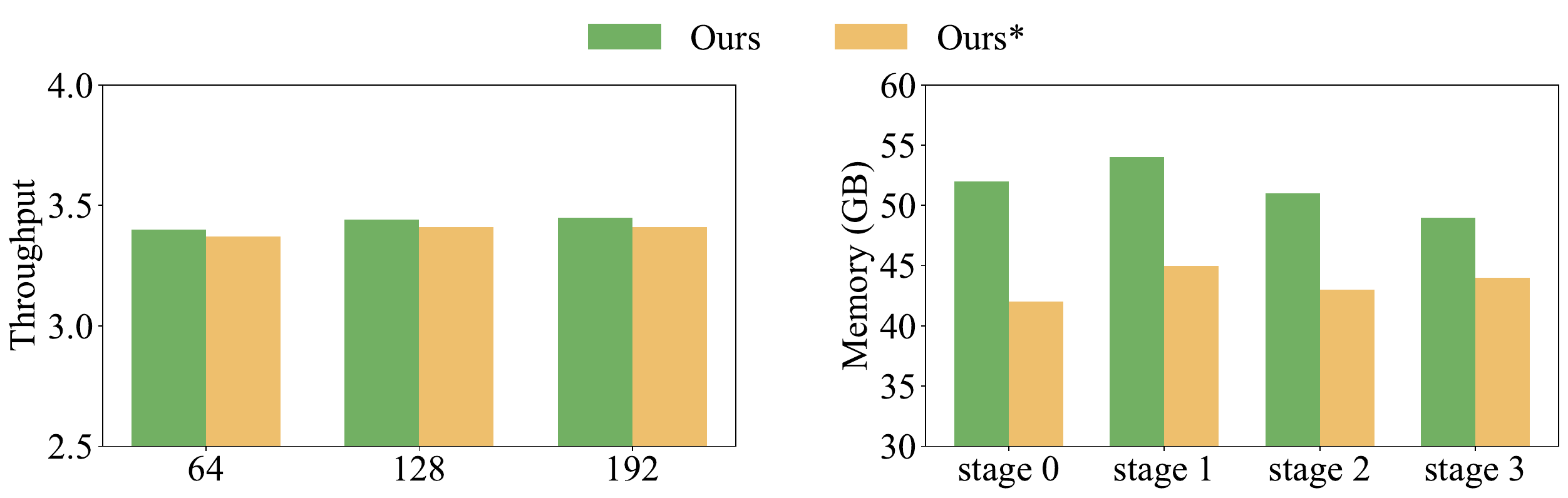}
    %%\vspace{-2mm}
    \caption{Experimental results in terms of throughput and peak activation memory footprints over 4 PP stages on the 12.1B LLM. $*$ indicates the enhanced variant with offloading.}
    \label{fig:offload result}
\end{figure}

\subsection{Efficiency of Our Enhanced Schedule}
\label{expe}

Due to the limited PCIe bandwidth of the A800, where the effectiveness of the offloading variant cannot be fully demonstrated, we perform experiments on 16 NVIDIA H20 GPUs, which are equipped with PCIe Gen 5 interconnection. As shown in Figure~\ref{fig:offload result}, the offloading variant incurs a negligible throughput degradation due to additional CPU overheads associated with data transmission. However, as demonstrated in the right part of Figure~\ref{fig:offload result}, the memory footprints across all stages are approximately equal, indicating our schedule effectively utilizes the memory resources to improve the throughput. And the offloading strategy achieves a peak memory reduction ranging from 10\% to 19.2\%, with the resulting peak memory footprint being close to that of 1F1B-I (40G). 

\subsection{Results under Maximized Resource Utilization}
We conduct experiments on 16 NVIDIA H20 96GB GPUs, adjusting the microbatch size across various parallel configurations to maximize memory utilization and achieve optimal throughput and Model FLOPs Utilization (MFU). As shown in Table~\ref{tab:max memory}, our proposed method delivers superior performance under maximized memory utilization. Specifically, under the TP=2, PP=8 configuration, our enhanced schedule achieves slightly higher throughput (2.74 samples/s) and MFU (92.86\%) compared to 1F1B-I, despite the lower proportion of TP bubbles on H20 GPUs (see the Appendix~\ref{app:d} for details), while maintaining a lower and more balanced memory footprint of 68 GB.

\begin{table}[!htbp]
    \centering
    \caption{Experiment results on 12.1B LLM under maximized memory utilization, which are evaluated on 16 H20 GPUs. $*$ denotes the enhanced variant of our schedule with activation offloading.}
    \vspace{2mm}
    \label{tab:max memory}
    \begin{tabular}{ccccccc}
    \toprule
    Configurations & Parallelism & $mbs$ & Schedules& Throughput & MFU & Memory  \\
    \midrule
    \multirow{24.5}{*}{\makecell[c]{$mbs$=192 \\ \\ $seq\_len$=8192}} & \multirow{4}{*}{\makecell[c]{TP=2 \\ pp=8}}& \multirow{4}{*}{1} & 1F1B-I & 2.72 & 92.09 & 76 \\
    & & & ZB-V & 2.61 & 88.36 & 54 \\
    & & & Ours & OOM \\
    & & & Ours* & \textbf{2.74} & \textbf{92.86} & 68 \\
    \cmidrule{2-7}
    & \multirow{6.5}{*}{\makecell[c]{TP=4 \\ \\ pp=4}}& \multirow{3}{*}{1} & 1F1B-I & 2.47 & 83.62 & 53 \\
    & & & ZB-V & 2.41 & 81.59 & 38 \\
    & & & Ours & 2.52 & 85.32 & 71 \\
    \cmidrule{3-7}
    & & \multirow{3}{*}{2} & 1F1B-I & OOM\\
    & & & ZB-V & 2.47 & 83.62 & 75 \\
    & & & Ours* & OOM \\
    \cmidrule{2-7}
    & \multirow{12}{*}{\makecell[c]{TP=8 \\ \\ pp=2}}& \multirow{3}{*}{1} & 1F1B-I & 2.06 & 69.74 & 40 \\
    & & & ZB-V & 2.07 & 70.08 & 31 \\
    & & & Ours & 2.12 & 71.78 & 57 \\
    \cmidrule{3-7}
    & & \multirow{3}{*}{2} & 1F1B-I & 2.14 & 72.45 & 76 \\
    & & & ZB-V & 2.18 & 73.81 & 61 \\
    & & & Ours* & OOM \\
    \cmidrule{3-7}
    & & \multirow{3}{*}{3} & 1F1B-I & OOM \\
    & & & ZB-V & 1.74 & 58.91 & 90 \\
    & & & Ours* & OOM \\
    \bottomrule
    \end{tabular}

\end{table}

\section{Conclusion}
In this paper, we presented a novel hybrid parallel strategy that synergistically integrates tensor and pipeline parallelism for distributed training of LLMs and MLLMs. By decomposing Transformer layers into fine-grained computation units and interleaving forward and backward computations within each PP stage, we constructed execution blocks that effectively overlap TP communication bubbles. Based on these blocks, we designed a ``V''-shape PP schedule,  which achieves balanced memory across stages and significantly reduces both TP and PP bubbles, albeit at the cost of increased peak memory. Experimental results demonstrated that our approach consistently outperforms existing scheduling methods on both LLMs and MLLMs. We also presented an enhanced variant incorporating offloading to better accommodate memory-constrained scenarios. In future work, we plan to explore more memory- and throughput-efficient hybrid schedules in the large-scale machine learning system. %%We also hope that this work can inspire further research in distributed community.

\section*{Acknowledgement}
This work was partly supported by the Funds for the National Natural Science Foundation of China under Grant 62202063, 62572072, U24B20176, and Beijing Natural Science Foundation (L243027).

{
    \small
    \bibliographystyle{plainnat}
    \bibliography{ref}
}

%%%%%%%%%%%%%%%%%%%%%%%%%%%%%%%%%%%%%%%%%%%%%%%%%%%%%%%%%%%%
% \newpage

%%%%%%%%%%%%%%%%%%%%%%%%%%%%%%%%%%%%%%%%%%%%%%%%%%%%%%%%%%%%

\newpage
\section*{NeurIPS Paper Checklist}

\begin{enumerate}

\item {\bf Claims}
    \item[] Question: Do the main claims made in the abstract and introduction accurately reflect the paper's contributions and scope?
    \item[] Answer: \answerYes{} % Replace by \answerYes{}, \answerNo{}, or \answerNA{}.
    \item[] Justification: We claim the contribution of our work clearly in the abstract and introduction.

\item {\bf Limitations}
    \item[] Question: Does the paper discuss the limitations of the work performed by the authors?
    \item[] Answer: \answerYes{} % Replace by \answerYes{}, \answerNo{}, or \answerNA{}.
    \item[] Justification: The main limitation of our work is discussed in the last section.

\item {\bf Theory assumptions and proofs}
    \item[] Question: For each theoretical result, does the paper provide the full set of assumptions and a complete (and correct) proof?
    \item[] Answer: \answerNA{} % Replace by \answerYes{}, \answerNo{}, or \answerNA{}.
    \item[] Justification: There is no  theoretical result in our work.

    \item {\bf Experimental result reproducibility}
    \item[] Question: Does the paper fully disclose all the information needed to reproduce the main experimental results of the paper to the extent that it affects the main claims and/or conclusions of the paper (regardless of whether the code and data are provided or not)?
    \item[] Answer: \answerYes{} % Replace by \answerYes{}, \answerNo{}, or \answerNA{}.
    \item[] Justification: Our implementation is based on open sourced M-Core, which is accessible from everyone and we describe the setting of experiments clearly.

\item {\bf Open access to data and code}
    \item[] Question: Does the paper provide open access to the data and code, with sufficient instructions to faithfully reproduce the main experimental results, as described in supplemental material?
    \item[] Answer: \answerYes{} % Replace by \answerYes{}, \answerNo{}, or \answerNA{}.
    \item[] Justification: : We would open the source code after the paper is accepted.

\item {\bf Experimental setting/details}
    \item[] Question: Does the paper specify all the training and test details (e.g., data splits, hyperparameters, how they were chosen, type of optimizer, etc.) necessary to understand the results?
    \item[] Answer: \answerYes{} % Replace by \answerYes{}, \answerNo{}, or \answerNA{}.
    \item[] Justification: Detailed in Section Experiments.

\item {\bf Experiment statistical significance}
    \item[] Question: Does the paper report error bars suitably and correctly defined or other appropriate information about the statistical significance of the experiments?
    \item[] Answer: \answerNo{} % Replace by \answerYes{}, \answerNo{}, or \answerNA{}.
    \item[] Justification: The results are stable in our clusters.

\item {\bf Experiments compute resources}
    \item[] Question: For each experiment, does the paper provide sufficient information on the computer resources (type of compute workers, memory, time of execution) needed to reproduce the experiments?
    \item[] Answer: \answerYes{} % Replace by \answerYes{}, \answerNo{}, or \answerNA{}.
    \item[] Justification: We list all resources information in Section Experiments.
    
\item {\bf Code of ethics}
    \item[] Question: Does the research conducted in the paper conform, in every respect, with the NeurIPS Code of Ethics \url{https://neurips.cc/public/EthicsGuidelines}?
    \item[] Answer: \answerYes{} % Replace by \answerYes{}, \answerNo{}, or \answerNA{}.
    \item[] Justification: The research is conducted with the NeurIPS Code of Ethics.

\item {\bf Broader impacts}
    \item[] Question: Does the paper discuss both potential positive societal impacts and negative societal impacts of the work performed?
    \item[] Answer: \answerNA{} % Replace by \answerYes{}, \answerNo{}, or \answerNA{}.
    \item[] Justification: Our research focus on the training of LLMs and has no impact on society.
    
\item {\bf Safeguards}
    \item[] Question: Does the paper describe safeguards that have been put in place for responsible release of data or models that have a high risk for misuse (e.g., pretrained language models, image generators, or scraped datasets)?
    \item[] Answer: \answerNA{} % Replace by \answerYes{}, \answerNo{}, or \answerNA{}.
    \item[] Justification: The paper poses no such risks.

\item {\bf Licenses for existing assets}
    \item[] Question: Are the creators or original owners of assets (e.g., code, data, models), used in the paper, properly credited and are the license and terms of use explicitly mentioned and properly respected?
    \item[] Answer: \answerYes{} % Replace by \answerYes{}, \answerNo{}, or \answerNA{}.
    \item[] Justification: Our work is developed based on Megatron-Core with a license that everyone can work on it.

\item {\bf New assets}
    \item[] Question: Are new assets introduced in the paper well documented and is the documentation provided alongside the assets?
    \item[] Answer: \answerYes{} % Replace by \answerYes{}, \answerNo{}, or \answerNA{}.
    \item[] Justification: We will open source the related documents and codes on the camera ready phase.

\item {\bf Crowdsourcing and research with human subjects}
    \item[] Question: For crowdsourcing experiments and research with human subjects, does the paper include the full text of instructions given to participants and screenshots, if applicable, as well as details about compensation (if any)? 
    \item[] Answer: \answerNA{} % Replace by \answerYes{}, \answerNo{}, or \answerNA{}.
    \item[] Justification: Not involve crowdsourcing nor research with human subjects.

\item {\bf Institutional review board (IRB) approvals or equivalent for research with human subjects}
    \item[] Question: Does the paper describe potential risks incurred by study participants, whether such risks were disclosed to the subjects, and whether Institutional Review Board (IRB) approvals (or an equivalent approval/review based on the requirements of your country or institution) were obtained?
    \item[] Answer: \answerNA{} % Replace by \answerYes{}, \answerNo{}, or \answerNA{}.
    \item[] Justification: Not involve crowdsourcing nor research with human subjects.

\item {\bf Declaration of LLM usage}
    \item[] Question: Does the paper describe the usage of LLMs if it is an important, original, or non-standard component of the core methods in this research? Note that if the LLM is used only for writing, editing, or formatting purposes and does not impact the core methodology, scientific rigorousness, or originality of the research, declaration is not required.
    %this research? 
    \item[] Answer: \answerNA{} % Replace by \answerYes{}, \answerNo{}, or \answerNA{}.
    \item[] Justification: The method in this research does not involve LLMs as any important, original, or non-standard components, just uses it for training tests.

\end{enumerate}

\newpage
\section*{Appendix}

\appendix
\section{The Warm-Up Phase of the Pipeline Schedule}
\begin{figure}[!ht]
    \centering
    \includegraphics[width=0.8\linewidth]{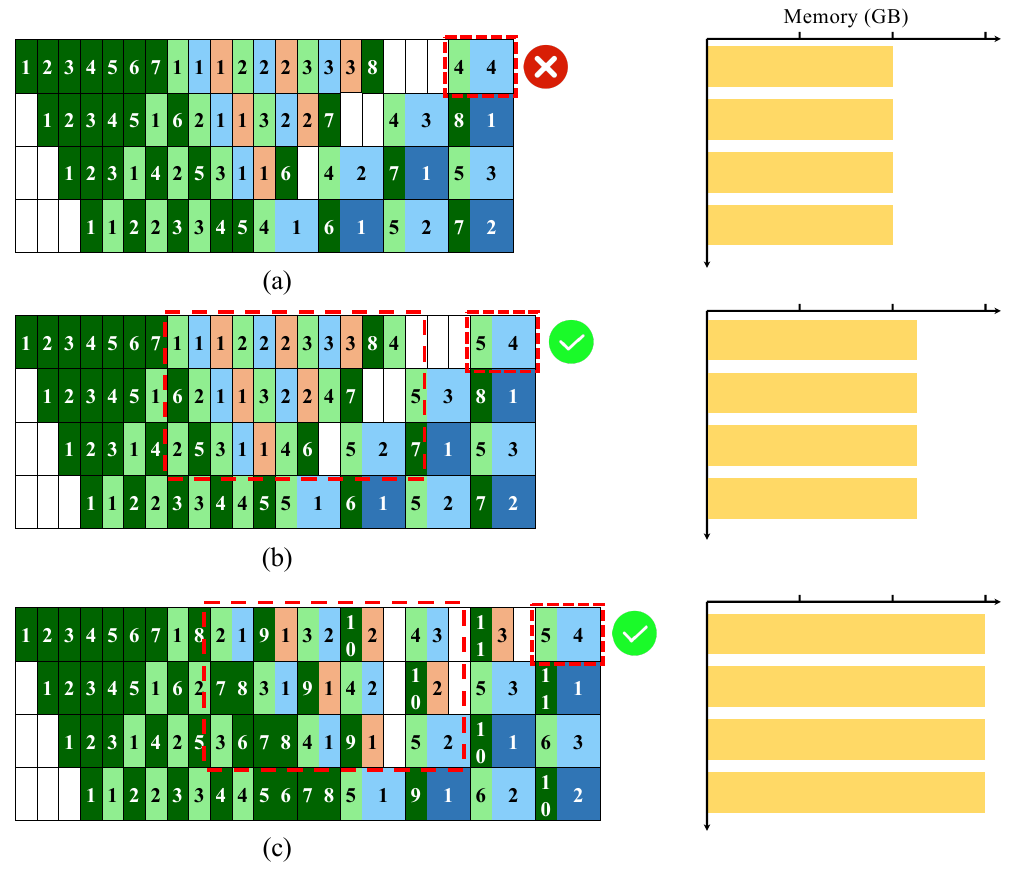}
    \caption{The construction of the warm-up phase: (a) wrong warm-up phase; (b) memory-efficient warm-up phase; (c) throughput-efficient warm-up phase.}
    \label{fig:blocks}
\end{figure}
The guiding principle for constructing the warm-up phase is to ensure the correct execution of overlapped forward and backward passes. As illustrated at the top of Figure~\ref{fig:blocks}, a key requirement is that overlapping between forward and backward passes must occur across different microbatches. Specifically, the microbatch index in the forward pass should be greater than that in the backward pass. To satisfy this condition, an additional forward pass is required before the overlapped F\&B execution begins, as shown in Figure~\ref{fig:blocks}(b). However, this approach necessitates decoupling the backward pass into activation and weight gradient computations, which exposes TP communication and introduces additional PP communication overheads. Please refer to Section~\ref{comp} for more details about the memory-efficient warm-up phase.

\section{Comparison with Other Schedules}
\label{comp}
\begin{figure}
    \centering
    \includegraphics[width=\linewidth]{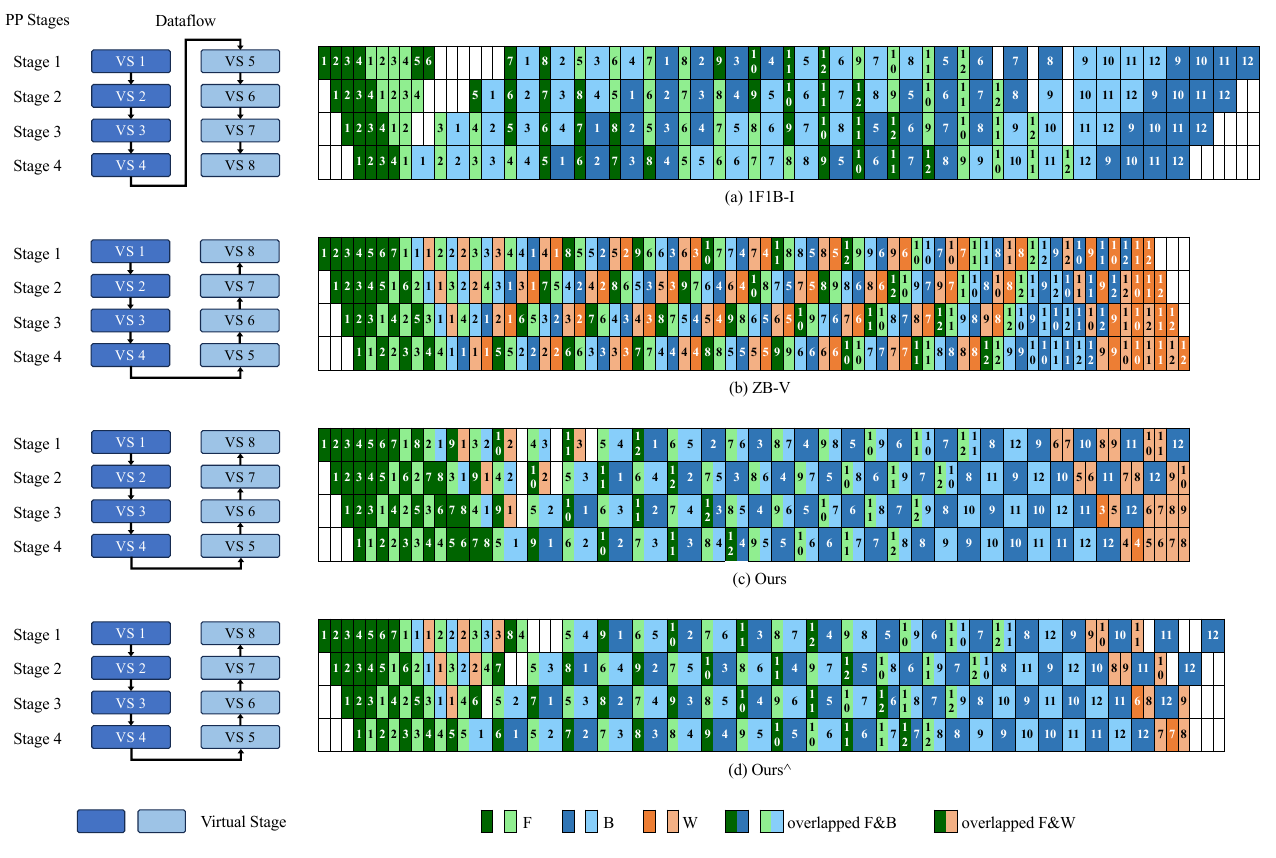}
    \caption{Synchronous pipeline parallelism schedules compared in our work, with 4 pipeline stages and 12 microbatches within a training iteration. Dark colors denote the first virtual stage and light colors are the second virtual stage. $\hat{}$ denotes the tensor and pipeline schedule beginning with the memory-efficient warm-up phase.}
    \label{fig:4sch}
\end{figure}

We compare 1F1B-I~\cite{narayanan2021efficient}, ZB-V~\cite{qi2024pipeline}, and our proposed schedule under identical configurations: 4 pipeline stages, 2 virtual stages per pipeline stage, and 12 microbatches, as illustrated in Figure~\ref{fig:4sch}. Intuitively, both ZB-V and our schedule have a similar number of PP bubbles, which are significantly fewer than those in 1F1B-I. The PP bubbles in 1F1B-I primarily result from the warm-up and cool-down phases. In the warm-up phase, the parallel dataflow across virtual stages results in a large number of bubbles. For instance, the long lifespan of virtual stage 2 for microbatch 1 cannot be efficiently filled without introducing additional forward passes from more microbatches, which would increases the peak memory of the stage 1. Consequently, 1F1B-I incurs greater PP bubble overheads during the warm-up phase. Meanwhile, the bubbles in the cool-down phase are inevitable due to inherent data dependencies.

Compared to ZB-V, our schedule results in a small number of PP bubbles at the end of the warm-up phase, which are attributed to the relatively long execution time and specific PP dataflow pattern of the overlapped F\&B blocks. In general, the activation gradient computation time $T_B$ satisfies $T_B > T_W$, where $T_W$ is the weight gradient computation time. As a result, small bubbles also appear at the end of the cool-down phase, and their magnitudes are approximately equal for both ZB-V and our proposed schedule.

Moreover, we also present schedule Ours$\hat{}$ (d), which begins with a memory-efficient warm-up phase, as illustrated in Figure~\ref{fig:4sch}. Although schedule (d) has a lower peak memory footprint compared to our standard schedule (c), it introduces additional PP bubbles towards the end, negatively impacting throughput performance. Therefore, for scenarios involving large-scale model training with large TP sizes and ample memory availability, we prefer schedule (c) for its higher performance. Our enhanced schedule variant, which includes activation offloading, is preferable for reducing peak memory usage, assuming sufficient bandwidth. Schedule (d) is suitable only for scenarios with extremely limited memory and large TP overheads.

\begin{table}[!ht]
\caption{Experimental results on 12.1B and 26.3B LLMs in terms of peak memory (GB), where the microbatch sizes of samples with sequence length 3072 and 2048 are set to 2.}
\label{app tab:memory}
\vspace{2.5mm}
\centering
\begin{tabular}{ccccccc}
\toprule
Model & \makecell[c]{Sequence \\ Length} & TP & PP & 1F1B-I & ZB-V & Ours\\
\midrule
\multirow{4.5}{*}{\makecell[c]{12.1B LLM \\ \\16 GPUs}} & \multirow{2}{*}{3072}& 4 & 4 & 41 & 30 & 53 \\
& & 8 & 2 & 31 & 24 & 43 \\
\cmidrule{2-7}
& \multirow{2}{*}{6144}& 4 & 4 & 41 & 30 & 54\\
& & 8 & 2 & 31 & 24 & 43 \\
\midrule
\multirow{4.5}{*}{\makecell[c]{26.3B LLM \\ \\32 GPUs}} & \multirow{2}{*}{2048}& 4 & 8 & 55 & 38 & 72 \\
& & 8 & 4 & 43 & 32 & 59 \\
\cmidrule{2-7}
& \multirow{2}{*}{4096}& 4 & 8 & 55 & 38 & 72\\
& & 8 & 4 & 43 & 32 & 59 \\
\bottomrule
\end{tabular}
\end{table}

\section{Detailed Experimental Results on LLMs}
We present detailed experimental results based on the 12.1B and 26.3B Qwen2 models, as summarized in Tables~\ref{app tab:memory}, \ref{tab:lm 2n}, and~\ref{tab:lm 2n mfu}. Notably, the GPU memory footprint reported in Table~\ref{app tab:memory} is slightly higher than the theoretical estimate. We attribute this discrepancy primarily to practical aspects of our engineering implementation, such as the implementation of the fine-grained computation units and the decoupling of weight gradient computation. For example, under TP=8, profiling the activation memory of one microbatch per virtual stage shows that ZB-V consumes 3.6 GB, while our approach uses 4.3 GB, an increase of nearly 20\%. Although this issue has no impact on throughput, we plan to further refine the engineering implementation to reduce this overhead in future work.

\begin{table}[!ht]
\centering
\caption{Experimental results on 12.1B and 26.3B LLMs in terms of throughput (samples per second). $mbs$ represents the number of microbatches.}
\label{tab:lm 2n}
\vspace{2.5mm}
\begin{tabular}{cccccccc}
\toprule
Model & \makecell[c]{Sequence \\ Length} & TP & PP & $mbs$ & 1F1B-I & ZB-V & Ours \\
\midrule
\multirow{13.5}{*}{\makecell[c]{12.1B LLM \\ \\16 GPUs}} & \multirow{6.5}{*}{3072} & \multirow{3}{*}{4} & \multirow{3}{*}{4} & 64 & 9.52 & 9.12 & 9.87 \\
 & & & & 128 & 9.63 & 9.26 & \textbf{10.1} \\
 & & & & 192 & 9.66 & 9.31 & \textbf{10.1} \\
 \cmidrule{3-8}
 & & \multirow{3}{*}{8} & \multirow{3}{*}{2} & 64 & 6.57 & 6.42 & 7.28 \\
 & & & & 128 & 6.60 & 6.46 & 7.32 \\
 & & & & 192 & 6.60 & 6.46 & \textbf{7.33} \\
 \cmidrule{2-8}
 & \multirow{6.5}{*}{6144} & \multirow{3}{*}{4} & \multirow{3}{*}{4} & 64 & 4.51 & 4.45 & 4.74 \\
 & & & & 128 & 4.57 & 4.48 & 4.82 \\
 & & & & 192 & 4.58 & 4.49 & \textbf{4.83} \\
 \cmidrule{3-8}
 & & \multirow{3}{*}{8} & \multirow{3}{*}{2} & 64 & 3.11 & 3.13 & 3.46 \\
 & & & & 128 & 3.13 & 3.13 & 3.47 \\
 & & & & 192 & 3.11 & 3.13 & \textbf{3.49} \\
\midrule
\multirow{13.5}{*}{\makecell[c]{26.3B LLM \\ \\32 GPUs}} & \multirow{6.5}{*}{2048} & \multirow{3}{*}{4} & \multirow{3}{*}{8} & 96 & 12.3 & 12.4 & 13.0 \\
 & & & & 176 & 12.8 & 12.7 & 13.2 \\
 & & & & 256 & 12.7 & 12.8 & \textbf{13.4} \\
 \cmidrule{3-8}
 & & \multirow{3}{*}{8} & \multirow{3}{*}{4} & 96 & 8.60 & 8.71 & 9.48 \\
 & & & & 176 & 8.67 & 8.79 & 9.56 \\
 & & & & 256 & 8.68 & 8.79 & \textbf{9.61} \\
 \cmidrule{2-8}
 & \multirow{6.5}{*}{4096} & \multirow{3}{*}{4} & \multirow{3}{*}{8} & 96 & 6.16 & 6.17 & 6.33 \\
 & & & & 176 & 6.17 & 6.28 & 6.49 \\
 & & & & 256 & 6.28 & 6.31 & \textbf{6.51} \\
 \cmidrule{3-8}
 & & \multirow{3}{*}{8} & \multirow{3}{*}{4} & 96 & 4.23 & 4.26 & 4.66 \\
 & & & & 176 & 4.24 & 4.28 & 4.70 \\
 & & & & 256 & 4.25 & 4.29 & \textbf{4.72} \\
\bottomrule
\end{tabular}
\end{table}

\begin{table}[!ht]
\centering
\caption{Experimental results on 12.1B and 26.3B LLMs in terms of Model FLOPs Utilization (MFU,~\%).}
\label{tab:lm 2n mfu}
\vspace{2.5mm}
\begin{tabular}{cccccccc}
\toprule
Model & \makecell[c]{Sequence \\ Length} & TP & PP & $mbs$ & 1F1B-I & ZB-V & Ours \\
\midrule
\multirow{13.5}{*}{\makecell[c]{12.1B LLM \\ \\16 GPUs}} & \multirow{6.5}{*}{3072} & \multirow{3}{*}{4} & \multirow{3}{*}{4} & 64 & 46.31 & 44.36 & 48.01 \\
 & & & & 128 & 46.84 & 45.04 & \textbf{49.13} \\
 & & & & 192 & 46.99 & 45.28 & \textbf{49.13} \\
 \cmidrule{3-8}
 & & \multirow{3}{*}{8} & \multirow{3}{*}{2} & 64 & 31.96 & 31.23 & 35.41 \\
 & & & & 128 & 32.10 & 31.42 & 35.61 \\
 & & & & 192 & 32.10 & 31.42 & \textbf{35.65} \\
 \cmidrule{2-8}
 & \multirow{6.5}{*}{6144} & \multirow{3}{*}{4} & \multirow{3}{*}{4} & 64 & 50.16 & 49.50 & 52.72 \\
 & & & & 128 & 50.83 & 49.83 & 53.61 \\
 & & & & 192 & 50.94 & 49.94 & \textbf{53.72} \\
 \cmidrule{3-8}
 & & \multirow{3}{*}{8} & \multirow{3}{*}{2} & 64 & 34.59 & 34.81 & 38.49 \\
 & & & & 128 & 34.81 & 34.81 & 38.60 \\
 & & & & 192 & 34.59 & 34.81 & \textbf{38.82} \\
\midrule
\multirow{13.5}{*}{\makecell[c]{26.3B LLM \\ \\32 GPUs}} & \multirow{6.5}{*}{2048} & \multirow{3}{*}{4} & \multirow{3}{*}{8} & 96 & 42.16 & 42.50 & 44.56 \\
 & & & & 176 & 43.87 & 43.53 & 45.24 \\
 & & & & 256 & 43.53 & 43.87 & \textbf{45.93} \\
 \cmidrule{3-8}
 & & \multirow{3}{*}{8} & \multirow{3}{*}{4} & 96 & 29.48 & 29.85 & 32.49 \\
 & & & & 176 & 29.72 & 30.13 & 32.77 \\
 & & & & 256 & 29.75 & 30.13 & \textbf{32.94} \\
 \cmidrule{2-8}
 & \multirow{6.5}{*}{4096} & \multirow{3}{*}{4} & \multirow{3}{*}{8} & 96 & 46.32 & 46.40 & 47.60 \\
 & & & & 176 & 46.40 & 47.23 & 48.80 \\
 & & & & 256 & 47.23 & 47.45 & \textbf{48.95} \\
 \cmidrule{3-8}
 & & \multirow{3}{*}{8} & \multirow{3}{*}{4} & 96 & 31.81 & 32.04 & 35.04 \\
 & & & & 176 & 31.88 & 32.19 & 35.34 \\
 & & & & 256 & 31.96 & 32.26 & \textbf{35.49} \\
 
\bottomrule
\end{tabular}
\end{table}

\section{Comparison with Other Schedule Methods on H20 GPUs}
\label{app:d}
\begin{figure}[!ht]
    \centering
    \includegraphics[width=0.8\linewidth]{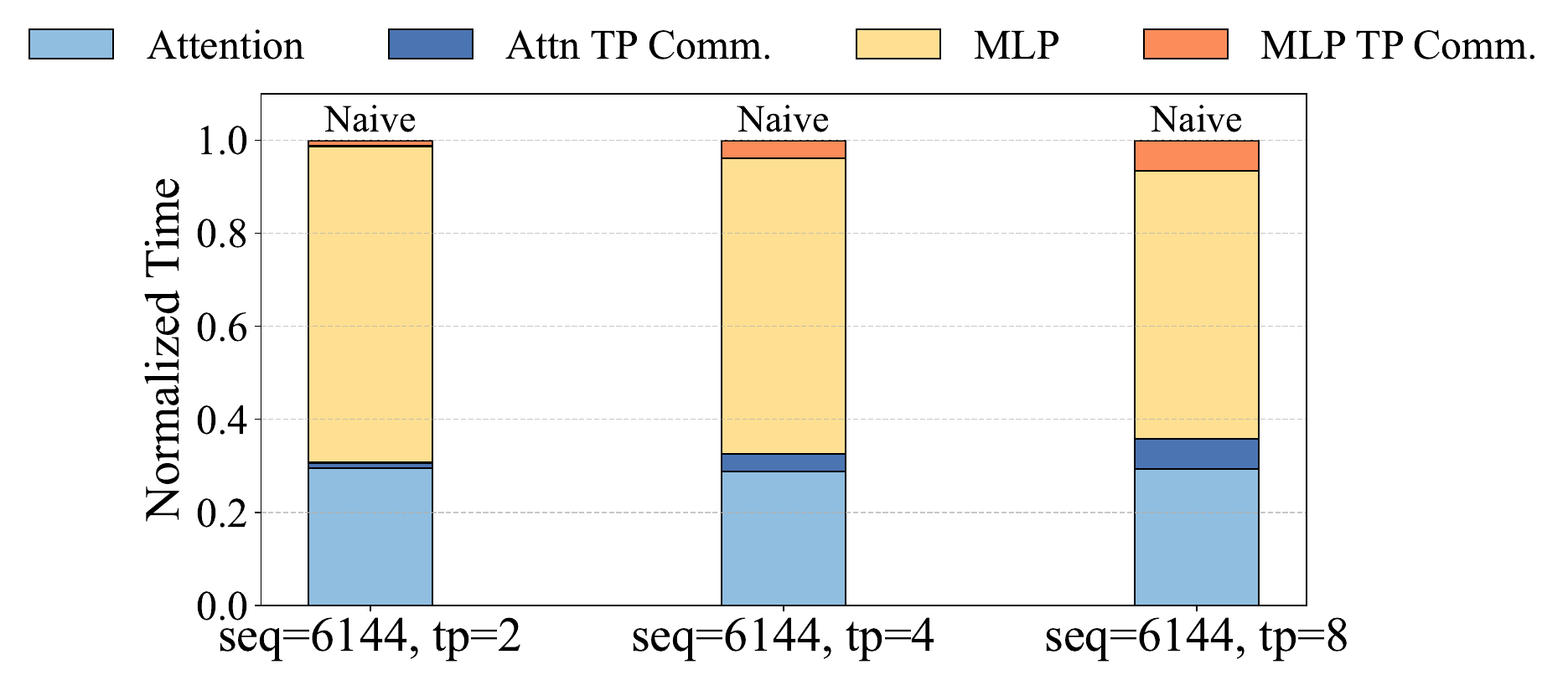}
    \caption{The computation and communication time proportion of Attention and MLP modules profiled on H20 GPUs.}
    \label{fig:TP ratio}
\end{figure}

We also conduct experiments on 16 NVIDIA H20 96G GPUs to evaluate the performance of our proposed schedule alongside other baselines, as summarized in Table~\ref{tab:h20}. Our proposed schedule outperforms other approaches by a margin, but the performance improvement is far less significant compared to that achieved on A800 GPUs. This discrepancy can be primarily attributed to the characteristics of the H20 GPUs, which have lower BF16 FLOPs but higher communication bandwidth. Furthermore, we profile the computation and communication times for the Attention and MLP modules of a single Transformer layer in the 12.1B Qwen2 model, as shown in Figure~\ref{fig:TP ratio}. It is evident that the proportion of TP communication is significantly lower than that measured on A800 GPUs. As a result, the benefits derived from optimizing TP bubbles have slightly diminished.

\begin{table}[!ht]
    \centering
    \caption{Experimental results evaluated on 12.1B LLM on 16 H20 GPUs. $mbs$ and $seq\_len$ denote the number of microbatches and the sequence length, respectively.}
    \label{tab:h20}
    \vspace{2.5mm}
    \begin{tabular}{ccccccc}
    \toprule
    Configurations & TP & PP & Schedules & Throughput & MFU & Memory  \\
    \midrule
    \multirow{10}{*}{\makecell[c]{$mbs$=192 \\ \\ $seq\_len$=6144}} & \multirow{3}{*}{2} & \multirow{3}{*}{8} & 1F1B-I & 3.77 & 88.34 & 58 \\
    & & & ZB-V & 3.63 & 85.06 & \textbf{42} \\ 
    & & & Ours & \textbf{3.79} & \textbf{88.81} & 73 \\ 
    \cmidrule{2-7}
    & \multirow{3}{*}{4} & \multirow{3}{*}{4} & 1F1B-I & 3.39 & 77.43 & 41 \\
    & & & ZB-V & 3.30 & 77.32 & \textbf{30} \\ 
    & & & Ours & \textbf{3.44} & \textbf{80.61} & 54 \\
    \cmidrule{2-7}
    & \multirow{3}{*}{8} & \multirow{3}{*}{2} & 1F1B-I & 2.80 & 65.61 & 31 \\
    & & & ZB-V & 2.80 & 65.61 & \textbf{24} \\ 
    & & & Ours & \textbf{2.89} & \textbf{67.72} & 43 \\ 
    \bottomrule
    \end{tabular}
\end{table}

\section{Evaluations with Existing Training Strategies}
\subsection{Activation Checkpointing}
Our proposed schedule is compatible with activation checkpointing strategies. We conducted the experiment on Qwen2-12.1B using activation checkpointing (AC) with a batch size of 128 and a sequence length of 6k, and report the results in the following table:

\begin{table}[!htbp]
\centering
\caption{Performance comparison with different activation checkpointing configurations.}
\label{tab:ac_comparison}
\vspace{2mm}
\begin{tabular}{lcc}
\toprule
\textbf{Config} & \textbf{Throughput (samples/s)} & \textbf{Peak Memory (GB)} \\
\midrule
AC disabled & 4.79 & 56.0 \\
AC enabled in Ours w/ MLP & 4.19 & 44.7 \\
AC enabled in Ours w/ Attn+MLP & 3.94 & 41.5 \\
AC enabled in Ours w/ Attn+MLP+Norm & 3.75 & 36.3 \\
\bottomrule
\end{tabular}
\end{table}

As shown in Table~\ref{tab:ac_comparison}, disabling AC yields the highest throughput of 4.79 samples per second but requires a peak memory footprint of 56.0 GB. Enabling AC selectively on the MLP modules reduces peak memory by 20.2\% to 44.7 GB, with a corresponding throughput decrease of 12.5\% to 4.19 samples/s. Extending AC to both Attention and MLP modules further reduces peak memory to 41.5 GB (a 25.9\% reduction), while throughput declines to 3.94 samples/s. The most aggressive configuration, applying AC to all modules, achieves the greatest memory savings, reducing peak memory by 35.2\% to 36.3 GB, at the cost of a 21.7\% reduction in throughput, resulting in 3.75 samples/s. These results confirm that our scheduling framework is fully compatible with activation checkpointing strategies.

\subsection{Data Parallelism and Context Parallelism}
\begin{table}[!ht]
    \centering
    \caption{Experimental results on 12.1B LLM compatible with data parallelism and context parallelism.}
    \vspace{2mm}
    \begin{tabular}{ccccccc}
    \toprule
    TP & PP & CP & $mbs$ & Seq & Schedules & Throughput \\
    \midrule
    \multirow{3}{*}{2} & \multirow{3}{*}{4} & \multirow{3}{*}{2} & \multirow{3}{*}{128} & \multirow{3}{*}{12k} & 1F1B-I & 2.64 \\
    & & & & & ZB-V & 2.61 \\
    & & & & & Ours & \textbf{2.71} \\
    \bottomrule
    \toprule
    TP & PP & DP & $mbs$ & Seq & Schedules & Throughput \\
    \midrule
    \multirow{3}{*}{2} & \multirow{3}{*}{4} & \multirow{3}{*}{2} & \multirow{3}{*}{256} & \multirow{3}{*}{4k} & 1F1B-I & 9.22 \\
    & & & & & ZB-V & 9.16 \\
    & & & & & Ours & \textbf{9.40} \\
    \bottomrule
    \end{tabular}
    \label{tab:cp}
\end{table}
To demonstrate the compatibility of our schedule with both data parallelism (DP) and context parallelism (CP), we conduct experiments on a 12.1B Qwen2 model. As shown in Table~\ref{tab:cp}, under identical tensor parallelism (TP=2) and pipeline parallelism (PP=4) configurations, our method achieves the highest throughput in both settings: 2.71 samples/s with CP (CP=2, $mbs=128$, sequence length 12k) and 9.40 samples/s with DP (DP=2, $mbs=256$, sequence length 4k). Our approach consistently outperforms baseline schedules (1F1B-I and ZB-V) across both parallelism paradigms, confirming its robustness and adaptability to different distributed training strategies.

\section{Impact of Communication Overlap on GEMM Execution Efficiency}
In distributed training of large language models, overlapping communication (e.g., AllReduce) with computation (e.g., GEMM) is widely used to improve hardware utilization and reduce training time. However, this overlap may cause resource contention, particularly for streaming multiprocessors (SMs) which potentially degrade GEMM performance.

To evaluate this effect, we conduct microbenchmarks under two scenarios: (1) GEMM dominates AllReduce, enabling full overlap; and (2) GEMM finishes early, leaving part of the communication exposed. In both cases, we compare overlapped execution against the sequential baseline (GEMM followed by AllReduce) and the standalone GEMM time.

\begin{table}[!ht]
    \centering
    \caption{Execution times (in milliseconds) for GEMM, AllReduce, and their combinations under sequential and overlapped execution.}
    \vspace{2mm}
    \begin{tabular}{lcc}
        \toprule
        Operation & Experiment 1 & Experiment 2 \\
        \midrule
        GEMM & 8.605 & 0.334 \\
        AllReduce & 3.364 & 1.643 \\
        GEMM + AllReduce (sequential) & 11.969 & 1.977 \\
        GEMM with overlapped AllReduce & 9.251 & 1.685 \\
        \bottomrule
    \end{tabular}
    \label{tab:gemm_ar_overlap}
\end{table}

The results, summarized in Table~\ref{tab:gemm_ar_overlap}, show that in the first scenario, where GEMM takes 8.605\,ms and AllReduce 3.364\,ms, the overlapped execution completes in 9.251\,ms, only 7.5\% slower than the GEMM alone and 22.6\% faster than the sequential execution (11.969\,ms). This indicates that the communication is effectively hidden with minimal interference.

In the second scenario, where GEMM is very short (0.334\,ms) and AllReduce dominates (1.643\,ms), the overlapped execution finishes in 1.685\,ms, which is just 2.6\% overhead over the AllReduce time and still 14.8\% faster than the sequential baseline (1.977\,ms). Notably, the GEMM itself is not measurably slowed down; rather, the slight increase in total time stems from the unavoidable tail of communication that cannot be hidden.

\end{document}